\documentclass[twocolumn,showpacs,pra]{revtex4}
\usepackage{amssymb}
\usepackage{graphicx}
\usepackage{dcolumn}
\usepackage{bm}
\usepackage{amsmath}
\usepackage{epsfig}

\begin{document}

\title{Deflection of field-free aligned molecules}

\author{E. Gershnabel}
\author{I. Sh. Averbukh}
\affiliation{Department of Chemical Physics, The Weizmann Institute
of Science, Rehovot 76100, ISRAEL}
\begin{abstract}

We consider deflection of  polarizable  molecules by inhomogeneous
optical fields, and analyze the role of molecular orientation and
rotation in the scattering process. It is shown that molecular
rotation induces spectacular rainbow-like features in the
distribution of the scattering angle. Moreover, by pre-shaping
molecular angular distribution with the help of short and strong
femtosecond laser pulses, one may efficiently control the scattering
process, manipulate the average deflection angle and its
distribution, and reduce substantially the angular dispersion of the
deflected molecules. This  opens new ways for many applications
involving molecular focusing, guiding and trapping by optical and
static fields.

\end{abstract}

\maketitle

Optical deflection of molecules by means of nonresonant laser fields
is a hot subject of many recent experimental studies
\cite{Deflection_general,Lens,Prism,Barker-new}. By controlling
molecular translational degrees of freedom \cite{Seideman}, novel
elements of molecular optics can be realized, including molecular
lens \cite{Deflection_general,Lens} and molecular prism
\cite{Prism}.  The mechanism of molecular deflection by a nonuniform
laser field is rather clear: the field induces molecular
polarization, interacts with it, and deflects the molecules along
the intensity gradient.  As most molecules have anisotropic
polarizability, the deflecting force depends on the molecular
orientation with respect to the deflecting field. Previous studies
on optical molecular deflection have mostly considered randomly
oriented molecules, for which the deflection angle is somehow
dispersed around the mean value determined by the
orientation-averaged polarizability. The latter   becomes
intensity-dependent for strong enough fields due to the
field-induced modification of the molecular angular motion
\cite{Zon,Friedrich}. This adds a new ingredient for controlling
molecular trajectories  \cite{Seideman,Barker-new}, which is
important, but somehow limited because of using the same fields for
the deflection process and orientation control.

 In this Letter, we show that the deflection process can be significantly affected and controlled
 by  \textit{pre-shaping} molecular angular distribution \emph{before} the molecules
enter the interaction zone.  This can be done with the help of
numerous recent techniques for laser molecular alignment, which use
single or multiple short laser pulses (transform-limited, or shaped)
to align molecular axes along certain directions.  Short laser
pulses excite rotational wavepackets, which results in a
considerable transient molecular alignment after the laser pulse is
over, i.e. at field-free conditions (for recent reviews on
field-free alignment, see, e.g. \cite{Stapelfeldt,Stapelfeldt1}).
Field-free alignment was observed both for small diatomic molecules
as well as for more complex molecules, for which full
three-dimensional control was realized \cite{3D1,3D2,3D3}. We
demonstrate that the average scattering angle of deflected molecules
and its distribution may be dramatically modified by a proper
field-free pre-alignment. By separating the processes of the angular
shaping and actual deflection, one gets a flexible tool for
tailoring molecular motion in inhomogeneous optical and static
fields.

 Although our arguments are rather general, we follow for certainty
a deflection scheme that reminds the experiment by Stapelfeldt $et$
$al$ \cite{Deflection_general} who used a strong IR laser to deflect
a $CS_2$ molecular beam, and then addressed a portion of the
deflected molecules (at a pre-selected place and time) by an
additional short and narrow ionizing pulse. Consider  deflection (in
$z$ direction) of a linear molecule moving in $x$ direction  with
velocity $v_x$ and interacting with a focused nonresonant laser beam
that propagates along the $y$ axis. The spatial profile of the laser
electric field in the $xz$-plane is $E=E_0\exp
[-(x^2+z^2)/\omega_0^2 ]\exp [-2\ln2t^2/\tau^2 ]$.  The interaction
potential of a linear molecule in the laser field is given by:
\begin{equation}
U(t)=-\frac{1}{4}E^2\left(\alpha_\parallel\cos^2\theta+\alpha_\perp\sin^2\theta\right),\label{U_general_eq}
\end{equation}
where $E$ is defined above, and $\alpha_\parallel$ and
$\alpha_\perp$ are the components of the molecular polarizability
along the molecular axis, and perpendicular to it, respectively.
Here $\theta$ is the angle between the electric field polarization
direction (along the laboratory $z$ axis) and the molecular axis. A
molecule initially moving along the $x$ direction will acquire a
velocity component $v_z$ along $z$-direction. We consider the
perturbation regime (weak field approximation) corresponding to a
small deflection angle, $\gamma\thickapprox v_z/v_x$. We substitute
$x=v_x t$,  and consider $z$ as a fixed impact parameter. The
deflection velocity is given by:

\begin{equation}\label{Velocity_Deflection}
v_z = \frac{1}{M}\int_{-\infty}^{\infty}F_z(t)dt
=-\frac{1}{M}\int_{-\infty}^{\infty}\left(\overrightarrow{\nabla}U(t)\right)_z,
\end{equation}

Here $M$ is the mass of the molecules, and $F_z$ is the deflecting
force. The time-dependence of the force $F_z(t)$ (and potential
$U(t)$) in Eq.(\ref{Velocity_Deflection}) comes from three sources:
pulse envelope, projectile motion of the molecule through the laser
focal area, and time variation of the angle $\theta$ due to
molecular rotation. For simplicity, we assume that the deflecting
field does not affect significantly the rotational motion. Such
approximation is justified, say for $CS_2$ molecules with the
rotational temperature $T=5K$, which are subject to the deflecting
field of $3\cdot10^9 W/cm^2$. The corresponding alignment potential
$U\approx-\frac{1}{4}\left(\alpha_\parallel-\alpha_\perp\right)E_0^2\approx
0.04\ meV$  is an order of magnitude smaller than the thermal energy
$k_BT$, where $k_B$ is Boltzmann's constant. This assumption is even
more valid if the molecules were additionally subject to the
aligning pulses prior to deflection.

  Since the rotational time scale is the shortest one in the problem,
 we average  the force over the fast rotation, and arrive at the following expression
 for the deflection angle, $\gamma = v_z/v_x$:
\begin{equation}\label{Deflection Angle}
\gamma = \gamma_0 \ \left[\alpha_{||}{\cal A} +\alpha_{\bot} (1 -{\cal A} )\right]/\overline{\alpha}
\end{equation}
Here $\overline{\alpha}=1/3 \alpha_{||}+2/3 \alpha_{\bot}$ is the orientation-averaged molecular polarizability, and
${\cal A}=\overline{\cos^2\theta}$ denotes the time-averaged
value of $\cos^2\theta$. This quantity depends on the relative
orientation of the vector of angular momentum and the polarization of
the deflecting field. It is different for different molecules of the
incident ensemble, which leads to the randomization of the
deflection process. The constant $\gamma_0$ presents the average deflection angle for an isotropic molecular ensemble:
\begin{eqnarray}\label{Average_Deflection Angle}
\gamma_0 &=&\frac{\overline{\alpha}E_0^2}{4Mv_x^2}\left(\frac{-4z}{\omega_0}\right) \nonumber\\
&\times&\sqrt{\frac{\pi}{2}}\left(1+\frac{2\omega_0^2\ln2}{\tau^2v_{x}^2}\right)^{-1/2}\exp\left(-\frac{2z^2}{\omega_0^2}\right)
\end{eqnarray}
We provide below some heuristic classical arguments on the
anticipated statistical properties  of ${\cal A}$ and  $\gamma$
(both for thermal and pre-aligned molecules) and then support them
by a more refined quantum treatment.

\begin{figure}[htb]
\begin{center}
\includegraphics[width=2cm]{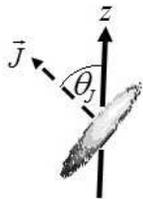}
\end{center}
\caption{A molecule rotates with a given angular momentum $\vec{J}$
that is randomly oriented in space. $\theta_J$ is the angle between
the angular momentum and the laboratory $z$ axis.}
\label{SimpleModel}
\end{figure}

Consider a linear molecule that rotates freely in a plane that is
perpendicular to the vector $\overrightarrow{J}$ of the angular
momentum (see Fig.(\ref{SimpleModel})).
\\

The projection of the molecular axis on the vertical $z$-direction
is given by $\cos \theta(t)=  \cos(\omega t) \sin\theta_J$, where
$\theta_J$ is the angle between $\vec{J}$ and $z$-axis, and $\omega$
is the angular frequency of molecular rotation. Averaging over time,
one arrives at:
\begin{equation}
{\cal A}=\overline{\cos^2\theta}=\frac{1}{2}
\sin^2\theta_J.\label{Simple_COS_avg_time}
\end{equation}
In a \emph{thermal} ensemble, vector $\vec{J}$ is randomly oriented
in space, with  isotropic angular distribution $1/2 \sin(\theta_J
)d\theta_J$.  The mean value of the deflection angle is $\langle
\gamma \rangle=\gamma_0$.  Eq.(\ref{Simple_COS_avg_time}) allows us
to obtain the distribution function, $f({\cal A})$ for ${\cal A}$
(and the related deflection angle) from the known isotropic
distribution for $\theta_J$. Since the inverse function
$\theta_J({\cal A})$ is multivalued, one obtains
\begin{equation}
f({\cal A})=\sum_{i=1}^2 \frac{1}{2} \sin\theta_J^{(i)}{\left|
\frac{d{\cal
A}}{d\theta_J^{(i)}}\right|}^{-1}=\frac{1}{\sqrt{1-2{\cal
A}}},\label{Gamma_Dist}
\end{equation}
where we summed over the two branches of $\theta_J({\cal A})$. This
formula predicts an \emph{unimodal  rainbow} singularity in the
distribution of the scattering angles at the maximal value
$\gamma=\gamma_0 (\alpha_{||} +\alpha_{\bot})/2\overline{\alpha}$ \
(for ${\cal A}=1/2$), and a flat step near the minimal one
$\gamma=\gamma_0 \alpha_{\bot}/\overline{\alpha}$ \ (for ${\cal
A}=0$). Assume now that the molecules are pre-aligned before
entering the deflection zone by a strong and short  laser pulse that
is polarized \emph{perpendicular} to the polarization direction of
the deflecting field (e.g., in $x$-direction). Such a pulse forces
the molecules to rotate preferentially in the planes containing the
$x$-axis. As a result, the vector $\vec{J}$ of the angular momentum
is confined to the $yz$-plane, and angle $\theta_J$ becomes
uniformly distributed in the interval $[0,\pi]$ with probability
density $d\theta_J/\pi$.
 The corresponding probability distribution for ${\cal A}$  takes the form
\begin{equation}
f({\cal A})=\frac{\sqrt{2}}{\pi}\frac{1}{\sqrt{{{\cal A}(1-2{\cal A}})}} \label{Gamma_Dist1}
\end{equation}
In contrast to Eq.(\ref{Gamma_Dist}), formula Eq.(\ref{Gamma_Dist1})
suggests a \emph{bimodal rainbow} in the distribution of deflection
angles, with singularities both at the minimal and the maximal
angles. Finally, we proceed to the most interesting case when the
molecules are pre-aligned by a short strong laser pulse that is
polarized \emph{parallel} to the direction of the deflecting field.
After excitation by such a pulse, the vector of the angular momentum
of the molecules is preferentially confined to the $xy$-plane, and
angle $\theta_J$ takes a well defined value of
$\theta_J\approx\pi/2$. As a result, the dispersion of the
scattering angles is  reduced dramatically. The distribution of the
deflection angle $\gamma$
  transforms to a narrow peak (asymptotically - a
$\delta$-function) near the maximal value, $\gamma=\gamma_0
(\alpha_{||} +\alpha_{\bot})/2\overline{\alpha}$.

\begin{figure}[htb]
\begin{center}
\includegraphics[width=8cm]{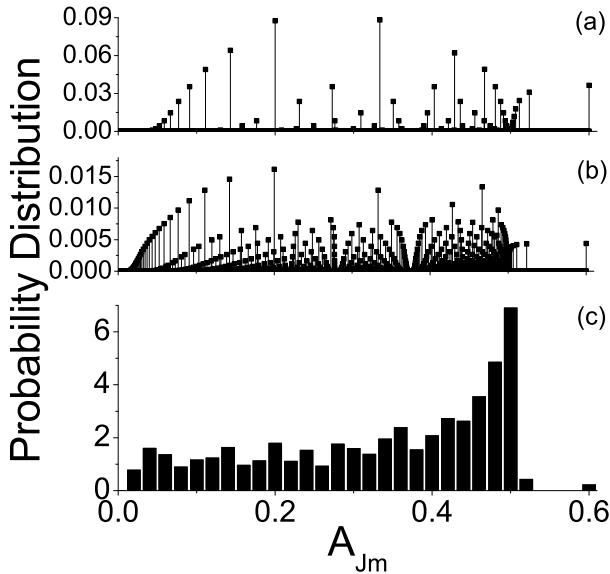}
\end{center}
\caption{Quantum distribution of ${\cal A}_{J,m}$ in the thermal
case. Panels (a) and (b) correspond to   $J_T=5$ and $J_T=15$,
respectively. Histogram in panel (c) presents a coarse-grained
continuous normalized distribution of ${\cal A}_{J,m}$ produced from
(b) by averaging over a set of finite bins.}\label{Quant thermal}
\end{figure}

For a more quantitative treatment, involving analysis of the
relative role of the quantum and thermal effects on one hand, and
the strength of the pre-aligning pulses on the other hand, we
consider quantum-mechanically the deflection of a linear molecule
described by the Hamiltonian ${\cal H}= \hat{J}^2/(2I)$. Here
$\hat{J}$ is operator of angular momentum, and $I$ is the moment of
inertia, which is related to the molecular rotational constant,
$B=\hbar/(4\pi Ic)$  ($c$ is speed of light). Assuming again that
the deflecting field is too weak to modify molecular alignment, we
consider scattering in different $|J,m\rangle$ states independently.
The deflection angle is given by Eq.(\ref{Deflection Angle}), in
which ${\cal A}$ is replaced by
\begin{equation}
{\cal A}_{J,m}=\langle J,m|\cos^2\theta |J,m\rangle
=\frac{1}{3}+\frac{2}{3}\frac{J(J+1)-3m^2}{(2 J + 3) (2 J - 1)}.
\label{AJm}
\end{equation}
In the quantum case, the continuous distribution of the angles
$\gamma$ is replaced by a set of discrete lines, each of them
weighted by the population of the state $|J,m\rangle$. Fig.
\ref{Quant thermal} shows the distribution of ${\cal A}_{J,m}$ in
the thermal case for various values of the dimensionless parameter
$J_T=\sqrt{k_B T/(hBc)}$ that represents the typical "thermal" value
of $J$ (for $J_T\geq 1$).

The distribution of discrete values of ${\cal A}_{J,m}$ demonstrates
a non-trivial pattern, however it shows the expected unimodal
rainbow feature (see Eq.(\ref{Gamma_Dist}))  for large enough $J_T$
after the coarse-grained averaging .

If the molecules are subject to a strong femtosecond pre-aligning
pulse, the corresponding interaction potential is given by
Eq.(\ref{U_general_eq}), in which $E(t)$ is replaced by the envelope
$\epsilon (t)$ of the femtosecond pulse.  If the pulse is short
compared to the typical periods of molecular rotation, it may be
considered as a delta-pulse. In the impulsive approximation, one
obtains the following relation between the angular wavefunction
before and after the pulse applied at $t=0$:
\begin{equation}
\Psi(t=0^+)=\exp(iP\cos^2\theta)\Psi(t=0^-),\label{before_after}
\end{equation}
where the kick strength, $P$ is given by
$P=\left(1/4\hbar\right)\cdot
(\alpha_{||}-\alpha_{\bot})\int_{-\infty}^{\infty}\epsilon^2(t)dt$.
Here we assumed the vertical polarization (along $z$-axis) of the
pulse. Physically, the dimensionless kick strength, $P$ equals to
the typical amount of angular momentum (in the units of $\hbar$)
supplied by the pulse to the molecule. For the vertical polarization
of the laser field, $m$ is a conserved quantum number. This allows
us to consider the excitation of the states with different initial
$m$ values separately. In order to find $\Psi(t=0^+)$ for any
initial state, we introduce an artificial parameter $\xi$ that will
be assigned the value $\xi=1$ at the end of the calculations, and
define
\begin{equation}
\Psi_{\xi}=\exp\left[(iP\cos^2\theta)\xi\right]\Psi(t=0^-)
=\sum_{J}c_J(\xi)|J,m\rangle.\label{xi_relation}
\end{equation}
By differentiating both sides of Eq.(\ref{xi_relation}) with respect
to $\xi$, we obtain the following set of differential equations for
the coefficients $c_J$:
\begin{equation}
\dot{c}_{J'}=iP\sum_J c_J\langle
J',m|\cos^2\theta|J,m\rangle,\label{Differential equations}
\end{equation}
where $\dot{c}= dc / d\xi $. The diagonal matrix elements in
Eq.(\ref{Differential equations}) are given by Eq.(\ref{AJm}), the
off-diagonal ones can be found using recurrence relations for the
spherical harmonics \cite{Arfken}. Since $\Psi_{\xi=0}=\Psi(t=0^-)$
and $\Psi_{\xi=1}=\Psi(t=0^+)$ (see Eq.(\ref{xi_relation})), we
solve numerically this set of equations from $\xi=0$ to $\xi=1$, and
find $\Psi(t=0^+)$. In order to consider the effect of the
field-free alignment at thermal conditions, we repeated this
procedure for every initial $|J_0,m_0\rangle$ state. To find the
modified population of the $|J,m\rangle$ states, the corresponding
contributions from different initial states were summed together
weighted with the Boltzmann's statistical factors. For symmetric
molecules,  statistical spin factor should be taken into account.
For example, for $CS_2$ molecules in the ground electronic and
vibrational state, only even $J$ values are allowed due to the
permutation symmetry for the exchange of two Bosonic Sulfur atoms
(that have nuclear spin $0$).
\begin{figure}[htb]
\begin{center}
\includegraphics[width=8cm]{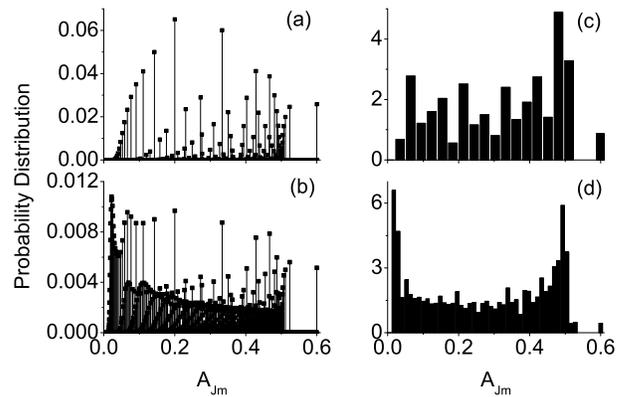}
\end{center}
\caption{Distribution of ${\cal A}_{J,m}$ for molecules pre-aligned
with the help of a short laser pulse polarized in the $x$ direction.
The left column (a-b) presents directly the ${\cal A}_{J,m}$ values,
while the right column (c-d) shows the corresponding coarse-grained
histograms (as in  Fig. \ref{Quant thermal}c). Panels (a) and (c)
are calculated for $J_T=5$ and $P=5$; (b) and (d) are for $J_T=5$
and $P=25$.} \label{DistributionT1P10}
\end{figure}
\begin{figure}[htb]
\begin{center}
\includegraphics[width=8cm]{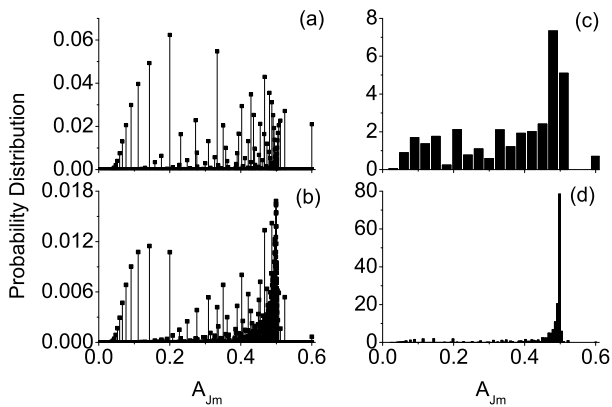}
\end{center}
\caption{Distribution of ${\cal A}_{J,m}$ for molecules pre-aligned
in the $z$ direction. The left column (a-b) presents directly the
${\cal A}_{J,m}$ values, while the right column (c-d) shows the
corresponding coarse-grained histograms. Panels (a) and (c) are
calculated for $J_T=5$ and $P=5$; (b) and (d) are for $J_T=5$ and
$P=25$.} \label{QuantDistP25T5}
\end{figure}
Using this technique, we considered deflection of initially thermal
molecules that were pre-aligned with the help of short pulses
polarized in $x$ and $z$ directions (Figs. \ref{DistributionT1P10}
and \ref{QuantDistP25T5}, respectively). In the case of the
alignment \emph{perpendicular} to the deflecting field, the
coarse-grained distribution of ${\cal A}_{J,m}$ (and that of the
deflection angles) exhibits the  \emph{bimodal rainbow} shape,
Eq.(\ref{Gamma_Dist1}) for strong enough kicks ($P\gg 1$ and $P\gg
J_T$). Finally, and most importantly, pre-alignment in the direction
\emph{parallel} to the deflecting field allows for almost complete
removal of the rotational broadening. A considerable narrowing of
the distribution can be seen when comparing Fig. \ref{Quant
thermal}a and Figs. \ref{QuantDistP25T5}b and \ref{QuantDistP25T5}d.

Our results indicate that pre-alignment provides an effective tool
for controlling the deflection of rotating molecules, and it may be
used for increasing the brightness of the scattered molecular beam.
This might be important for nano-fabrication schemes based on the
molecular optics approach \cite{Seideman}. Moreover, molecular
deflection by non-resonant optical dipole force is considered as a
promising route to separation of molecular mixtures (for a recent
review, see \cite{Chinese}). Narrowing  the distribution of the
scattering angles may substantially increase the efficiency of
separation of multi-component beams, especially when the
pre-alignment is applied selectively to certain molecular species,
such as isotopes \cite{isotopes}, or nuclear spin isomers
\cite{Fauchet,isomers}. More complicated techniques for pre-shaping
the molecular angular distribution may be considered, such as
confining molecular rotation to a certain plane by using the
"optical molecular centrifuge" approach \cite{Centrifuge},
double-pulse ignited "molecular propeller" \cite{NJP}, or
two-direction alignment alternation excited by elliptic laser pulses
\cite{alternation}. In this case, a narrow angular peak is expected
in molecular scattering, whose position is controllable by
inclination of the plane of rotation with respect to the deflecting
field. Laser pre-alignment may be used to manipulate molecular
deflection by inhomogeneous \emph{static} fields as well (for recent
exciting experiments on \emph{post-alignment} of molecules scattered
by static electric fields see \cite{post}). In particular, one may
affect molecular motion in relatively weak fields that are
insufficient to modify rotational states by themselves.  Moreover,
the same mechanisms may prove efficient for controlling inelastic
molecular scattering off metalic/dielectric surfaces. These and
other aspects of the present problem are subjects of an ongoing
investigation.

This research is made possible in part by the historic generosity of
the Harold Perlman Family. IA is an incumbent of the Patricia Elman
Bildner Professorial Chair.

\bibliographystyle{phaip}

\end{document}